\documentclass[12pt,openany]{article}
\usepackage{epsfig,longtable}

\begin{document}

\title 
      {$\Sigma^-$ Photoproduction on the Neutron:
      Results from CLAS}

\author{I. Niculescu \\
\\{\it The George Washington University, Washington, DC 20052}\\
for the CLAS Collaboration}  
\maketitle

\begin{abstract}
As part of a broader study of kaon photoproduction on deuterium, the
$\gamma n \to K^{+} \Sigma^{-} $ channel was investigated.
The data were acquired at Jefferson Lab using the CLAS detector and 
the Photon-Tagging Facility installed in Hall B.  
 The photon energy range covered was from 0.50 to 
2.95~GeV. For the present analysis, the $\gamma n \to K^{+} \Sigma^{-}$ channel
was identified by detecting the positive kaon in coincidence with both
decay products of the $\Sigma^-$ hyperon, $\pi^-$ and $n$. Preliminary differential 
cross--sections are shown as a function of the invariant energy W and the kaon
polar angle in the center of mass system.
\end{abstract}

\section{Introduction}

The electromagnetic strangeness production is an important part of the
Jefferson Lab's experimental program. Several experiments have been
approved to run in all three experimental halls.
 These experiments include kaon electro- and photoproduction
on hydrogen, deuterium, $^3$He, and $^4$He. 

Kaon photoproduction on deuterium is governed by three main
mechanisms:
\begin{itemize}
\item{} The elementary amplitudes of the six kaon production reactions possible
on the nucleon
  ($\gamma p \to K^{+} \Lambda $, $\gamma p \to K^{+} \Sigma^{0} $,
  $\gamma p \to K^{0} \Sigma^{+} $, $\gamma n \to K^{+} \Sigma^{-} $, 
  $\gamma n \to K^{0} \Lambda $, $\gamma n \to K^{0} \Sigma^{0} $).
\item{} The Fermi motion of the proton and neutron inside the
  deuteron. The momentum distribution of nucleon momenta can be
  calculated using the deuteron wave function.
\item{} The interaction between the final--state hadrons.   
\end{itemize}

\begin{figure}
   \begin{center}
    \leavevmode
    \epsfxsize=4.5in
    \epsffile{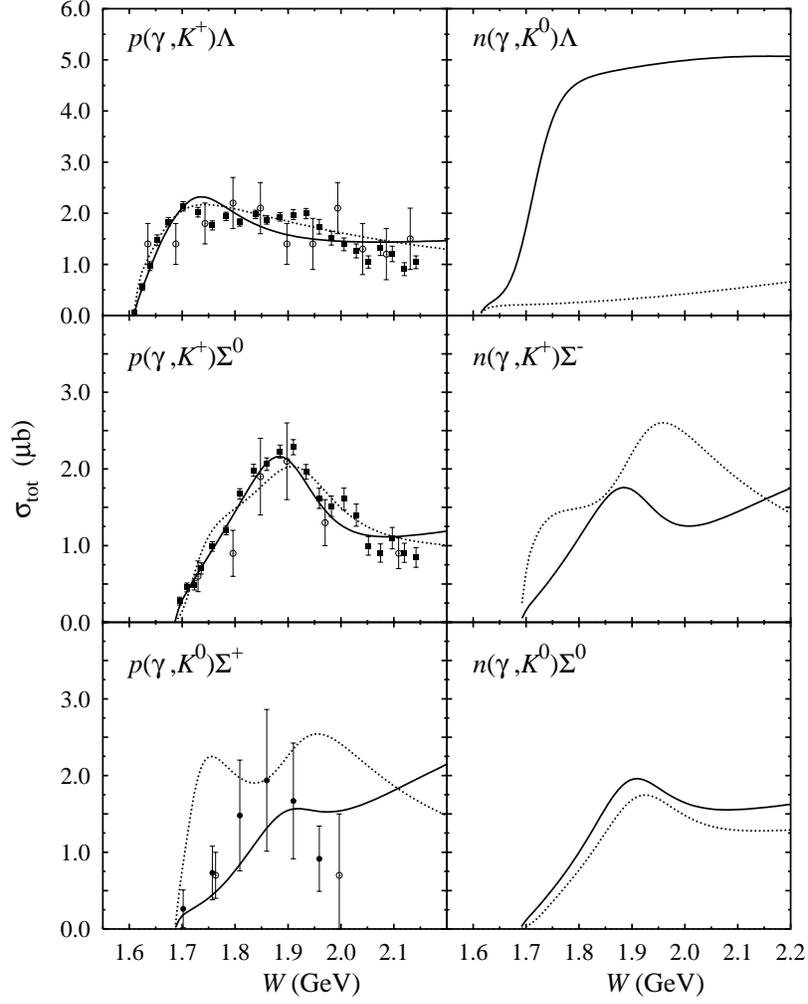}         
  \caption{The six possible kaon
      photoproduction channels compared
      to experimental data. The two curves represent two theoretical
      calculations by Yamamura {\it et al.} \protect \cite{1}.}
\label{fig:gamkelem}
\end{center}
\end{figure}

Experimental information exists for the first three exclusive kaon
photoproduction channels. As shown in Fig.\ \ref{fig:gamkelem}
\cite{1}, there are no previous 
data for the kaon photoproduction on the neutron. One of the goals of 
experiment E89-045 was to investigate open strangeness photoproduction on
the neutron. These studies will provide additional information about isospin 
dependence since the elementary operator for the $\Sigma^-$ production could be
quite different than the operator for $\Sigma^0$ production\ \cite{xili}.

\section{Experiment}

The data presented here were obtained in Hall B in 
August--September 1999. The total running time was 23 days. 
Real photons were produced by tagged 
bremsstrahlung provided by the Photon-Tagging Facility of Hall B
\cite{tagger}. 
Experimental data were obtained for two incident electron energies: 
2.47 and 3.11~GeV. Since the tagging range is 20\%~--~95\% of the 
electron-beam energy, the photon energy range covered was from 0.50 to 
2.95~GeV. The rate of tagged photons was approximately 10$^{7}$/sec. 
The outgoing particles were detected in the CEBAF Large Acceptance
Spectrometer (CLAS) \cite{2}, a magnetic toroidal multi--gap spectrometer,
covering a range of polar angles from $10^\circ$ to $150^\circ$, with almost $2 \pi$ azimuthal 
coverage.

\section{Data Analysis and Results}

\begin{figure}
   \begin{center}
    \leavevmode
    \epsfxsize=4.5in
    \epsffile{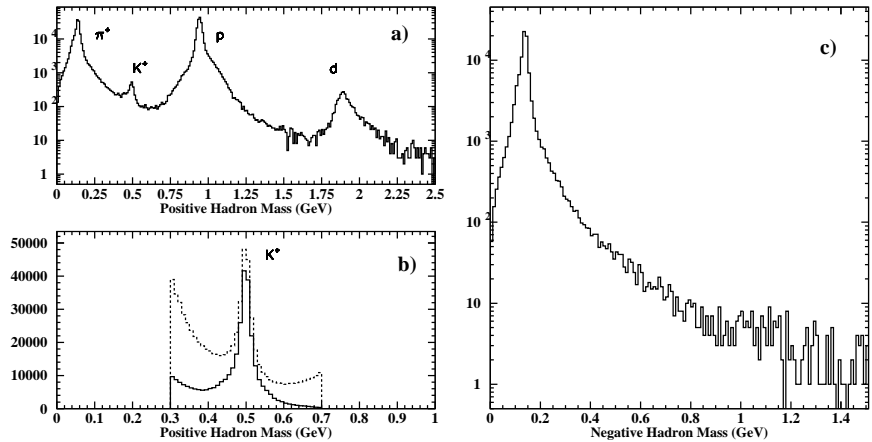}         
  \caption{a) The positive hadron mass as
      determined from time--of--flight measurement in CLAS. Panel b) 
shows the positive kaon mass before (dashed line) and after (solid line) 
applying additional timing cuts. c) The negative hadron mass as
      determined from time--of--flight measurement in CLAS.}
\label{fig:kp_pid}
\end{center}
\end{figure}

Charged particle identification is achieved in CLAS using information
about the flight time and momentum of the particle obtained using the
time--of--flight (TOF) detectors and drift chambers, respectively.
 A spectrum for positive hadrons as detected in
CLAS is shown in panel a) of Figure~\ \ref{fig:kp_pid}. The
peak corresponding to the positive kaons is barely visible between the
pion and proton peaks. The dashed curve in the bottom panel shows the 
kaon peak after filtering the data by applying a mass cut from 0.3 to 0.7.  
There is still a significant background underneath the kaon peak,
coming mostly from misidentified pions and protons.
The solid curve represents the same spectrum after applying additional timing 
cuts to minimize the number of pions and protons produced by photons coming 
from neighbouring beam bunches. 

In the present analysis the reaction $\gamma n \to K^+\Sigma^-$ was
selected by detecting the positive kaon and the decay products of
the $\Sigma^-$, the neutron and the negative pion. The pion was
detected using the time--of--flight counters and the drift chambers 
(Figure\ \ref{fig:kp_pid} c)), while the
neutron was detected in the electromagnetic calorimeter. The neutron momentum
was determined from the time--of--flight information given by the 
electromagnetic calorimeter.

\begin{figure}
   \begin{center}
    \leavevmode
    \epsfxsize=4.5in
    \epsffile{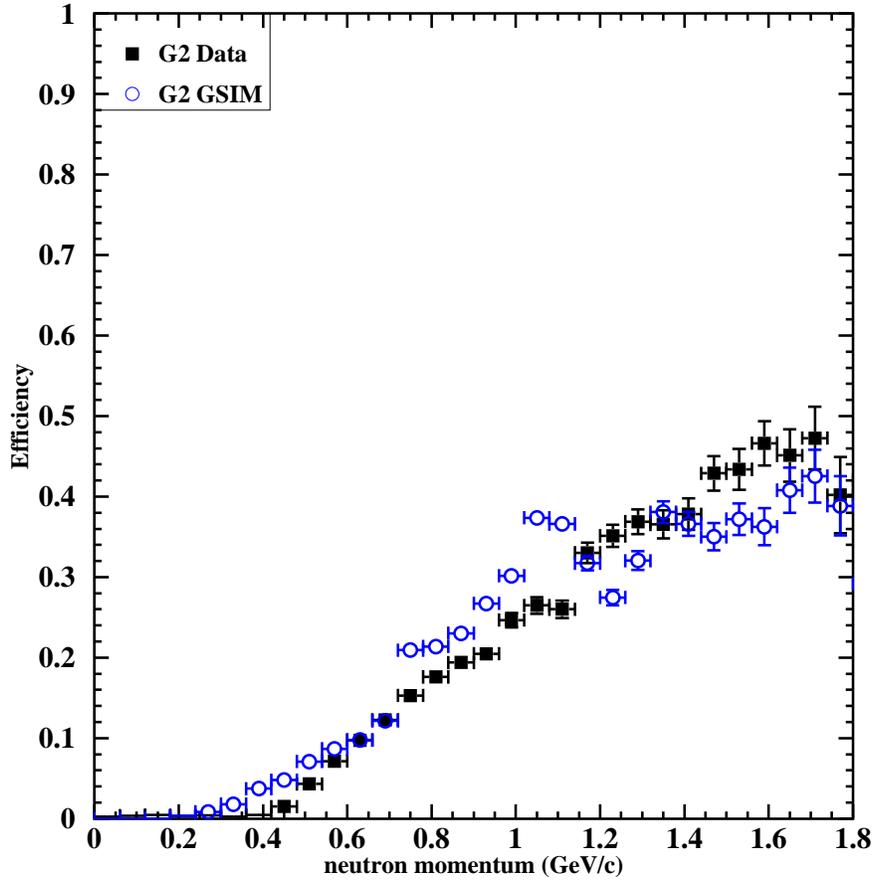}         
  \caption{Comparison of neutron detection
  efficiency from the data (full symbols) and Monte Carlo (open symbols) for 
the electromagnetic calorimeter in CLAS}
\label{fig:n_pid}
\end{center}
\end{figure}

The efficiency for detecting neutrons in CLAS was studied using the
reaction $\gamma d\to p n \pi^+ \pi^-$. The results  are shown in Figure\
\ref{fig:n_pid} compared to a GEANT--based Monte Carlo simulation of the CLAS 
detector (GSIM). The agreement between the data and Monte Carlo, while not 
perfect, was deemed reasonable for the limited precision required for the 
present experiment. Detection inefficiencies and geometric acceptance
corrections for charged particles were also obtained using GSIM. The events for the reaction 
$\gamma n \to K^+ 
\Sigma^-$ were generated as quasifree production on a neutron with
initial momentum distribution calculated using the Bonn potential\ 
\cite{Bonn,bonn2}. 

\begin{figure}   
\begin{center}
    \leavevmode
    \epsfxsize=4.5in
    \epsffile{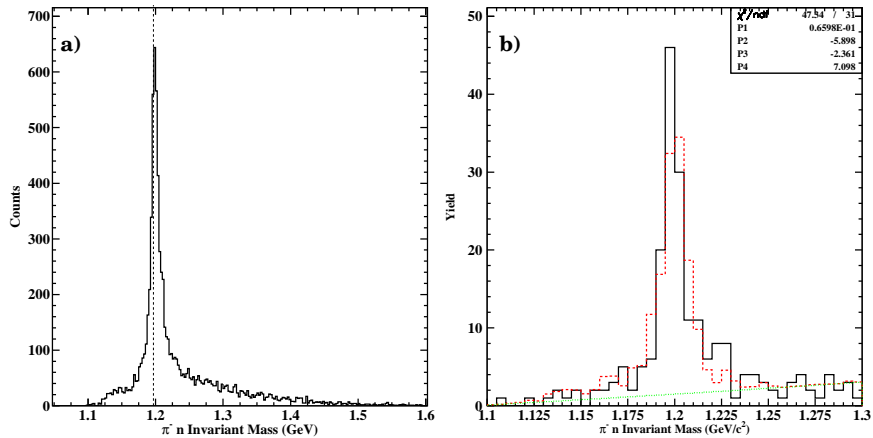}         
  \caption{a) $\Sigma^-$ Invariant mass, summed over all possible photon energies and all kaon angles
 b)$(\pi^-,n)$ invariant mass for one bin in photon beam energy 
 and $\theta_K^*$. The solid line represents the data, the dashed line 
shows the simulation plus the background, and the dotted line is the background.}
\label{fig:inv_mass}
\end{center}
\end{figure}

In Figure\ \ref{fig:inv_mass}a the invariant mass of the $(\pi^-,n)$ system 
reconstructed using the energy and momentum of the negative pion and neutron
is shown, integrated over all photon energies (incident electron beam energy
was 2.47~GeV) and $K^+$ angles. The expected value for the $\Sigma^-$ mass is 
indicated by the dashed vertical line. 

For the present analysis the data were binned in a 2D grid with 100~MeV bins in incident 
photon energy and five bins in the center of mass polar angle of the $K^+$ ($\theta_{K}^*$). 
For each bin a cut around the $\Sigma^-$ peak
was used to identify the $\gamma + n\to K^+ + \Sigma^-$ events.  
Due to the large uncertainty in the determination of the neutron momentum,
this cut was kept loose, so as to minimize the rejection of good $\Sigma^-$ 
events. 

The experimental yield was extracted by fitting ( using a using
a log-likelihood method) the shape of the invariant mass, as obtained 
from the simulation to the experimental data, allowing for a small ammount 
of residual background.
This procedure is illustrated in Figure\ \ref{fig:inv_mass}b for events having photon energies between 
1.4 and 1.5~GeV and $0.1 \le \cos (\theta_{K}^*) \le 0.4$: 
The dashed line shows the
simulation (signal) plus the background, while the dotted line represents the background alone.
 The data are represented by the solid line.  
 
In order to obtain the differential cross--section the extracted yield was corrected 
for all known detector inefficiencies and geometric acceptance effects. The target 
density and length as well as the tagged photon 
flux were also taken into account\ \footnote{Note that at present time there is 
a $\sim$20\% uncertainty in the determination of the photon flux. Final normalization 
constants will be available in the near future.}. 

\begin{figure}
\begin{center}
    \leavevmode
    \epsfxsize=4.5in
    \epsffile{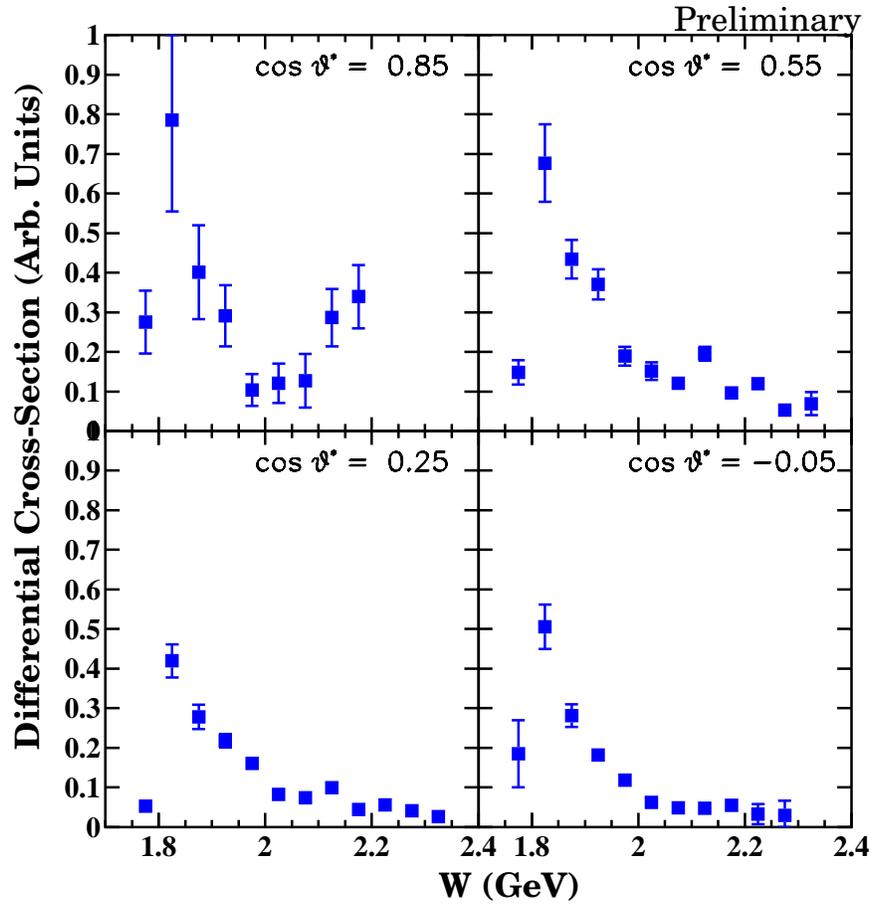}         
  \caption{Preliminary differential cross--sections for $\gamma n \to K^+ \Sigma^-$ as a 
function of W for four bins in the kaon (center of mass) angle.}
\label{fig:myresults}
\end{center}
\end{figure}

Figure\ \ref{fig:myresults} shows preliminary differential cross--sections for four of 
the five angular bins. The uncertainties shown are statistical only. Extensive studies 
of the systematic uncertainties are well underway (current estimates indicate a systematic 
uncertainty of 15\% or less for most bins).

For all angles one sees that the cross--section rises rather sharply from threshold, 
reaching a maximum for W between 1.8 and 1.9~GeV. The data do not rule out the existence 
of structures (bumps) in the higher W range, although further investigation is needed.

\section{Summary}

The reaction $\gamma d \to K^+ \Sigma^- (p)$ was studied using tagged photons and the 
CLAS detector in Hall B at Jefferson Lab. The strange $K^+$ meson was detected before 
its in--flight decay, while the associated $\Sigma^-$ hyperon was identified by observing 
both its decay products ($\pi^-$ and $n$). Preliminary differential cross--sections 
were presented. The current analysis provides complementary information to that obtained 
in photoproduction off the proton, furthering our understanding of the reaction mechanism 
that governs open strangeness production. Future work, besides finalizing the present 
analysis, will focus on the extraction of the final state YN interaction.

\section{Acknowledgments}
This work was supported in part by the U.~S.~Department of Energy under 
Grant No. DE-FG02-95ER40901. 

\bibliography{}

\end{document}